\documentclass[12pt]{article} 
\usepackage[koi8-r]{inputenc}
\usepackage{cyr}
\usepackage{epsf}
\usepackage{pazha}
\usepackage[round]{natbib}
\usepackage{graphicx}
\usepackage{hyperref}
\tightenlines

\newcommand{\degree}{^\circ}

\usepackage{amsmath} 
\usepackage{xcolor}

\sloppypar

\begin{document}
\baselineskip 21pt


\title{\bf Simulation of the Free Fall of a Gas Stream
on a Protoplanetary Disk}

\author{\bf \hspace{-1.3cm}\copyright\, 2024 \ \ 
 V. V. Grigoryev, T. V. Demidova,}

\affil{
{\it Crimean Astrophysical Observatory of the Russian Academy of Sciences,\\ Nauchny, Crimea, Russia}}

\received{18.07.2024;
Received March 20, 2024; revised June 12, 2024;\\ accepted July 18, 2024}
\noindent
\begin{small}
\textbf{Ast. Zhur., 2024, tom 101, No 10, pp. 866-884. DOI: 10.31857/S0004629924100014}\\
\textbf{Ast. Rep., 2024, Vol. 68, No. 10, pp. 949-966. DOI: 10.1134/S1063772924700859}
\end{small}

\sloppypar 
\vspace{2mm}
The problem of the formation of exoplanets in inclined orbits relative to the equatorial plane of the parent star or the main plane of the protoplanetary disk can be solved by introducing a smaller inclined disk. However, the question of the nature of such an internal disk remains open. In the paper, we successfully tested the hypothesis about the formation of an inclined inner disk in a protoplanetary disk near a T Tau type star as a result of a gas stream falling on it. To test the hypothesis, three-dimensional gas-dynamic calculations
were performed taking into account viscosity and thermal conductivity using the PLUTO package. In the course of the analysis of calculations, it was shown that a single intersection of the matter stream with the plane of the disk cannot ensure the formation of an inclined disk near the star, while a double intersection
can. In addition, in the case of a retrograde fall of matter, the angle of inclination of the resulting inner disk is significantly greater. An analysis of the observational manifestations of this event was also carried out: the potential change in the brightness of the star, the distribution of optical thickness in angles, the evolution of the accretion rate. It is shown that the decrease in brightness can reach up to $5^m$, taking into account scattered light, and such a decrease in brightness will last several decades. In addition, a sharp increase in the accretion rate by two orders of magnitude could potentially trigger an FU Ori-like outburst.



\vfill
\noindent\rule{8cm}{1pt}\\
{$^*$ email: vitaliygrigoryev@crao.ru}

\clearpage

\section{INTRODUCTION}
\noindent

The discovery of a tilt in the inner part of the debris
disk of $\beta$ Pic~\citep{1995AAS...187.3205B} stimulated the development of theories explaining the formation of similar irregularities in the protoplanetary disks of young stars. It was shown in~\citet{1997MNRAS.292..896M,1997MNRAS.285..288L} that the tilt of the inner parts of the disk may be a consequence of the motion of a low-mass companion on an orbit inclined relative to the plane of the
disk. Subsequently, the planet itself was discovered~\citep{2009A&A...506..927L,2012A&A...542A..41C} at a distance of 8--15 AU. In case of $\beta$ Pic the orbital inclination is several degrees.

A similar picture was discovered in the disk of the star CQ Tau (a representative of the Herbig Ae class), the inner part of the disk of which is tilted by $\sim 30\degree$ relative to the outer one~\citep{2004ApJ...613.1049E,2008A&A...488..565C}.

The low-mass star AA Tau was a variable star with brightness variations with a period of 8.5 days, which is close to the star's rotation period~\citep{1999A&A...349..619B}. The authors
attribute the star's variability to its periodic shielding
by an inclined inner disk, the existence of which is due
to the tilt of the magnetic dipole axis to the star's rotation axis. However, in 2011, the star's brightness dropped by $\sim 2^m$ in V band, which was accompanied by reddening in the near-IR region. The star has not yet returned to its bright state~\citep{2021AJ....161...61C}. An analysis of the AA~Tau image obtained in the ALMA (Atacama Large Millimeter Array) survey was performed in~\citet{2017ApJ...840...23L}. It showed that the protoplanetary disk of the star has a multi-band structure in millimeter dust. In addition, the inner disk ($\sim 10$~AU in size) is tilted relative to the disk periphery by $\sim 20\degree$. The authors also suggest that
there is a hump or a stream of matter on the disk surface that connects the inner and outer disks. An increase in the thickness of the inner disk or the passage of a stream of matter on the line of sight may be the cause of a significant drop in the star's brightness.

Observation of the Rossiter--McLaughlin effect~\citep{1924ApJ....60...15R,1924ApJ....60...22M} during transits of exoplanets across the disk of stars showed that for a significant number of objects
the orbital plane does not coincide with the equatorial
plane of the central star~\citep{2022PASP..134h2001A}. In addition to inclined orbits, there are also perpendicular (for example,
WASP-7~\citet{2009ApJ...690L..89H}) and retrograde orbits (for example, WASP-17~\citet{2010A&A...524A..25T}).

Simulation shown that if the line of sight crosses
the protoplanetary disk of a star with a low-mass companion on an inclined orbit, the light curve most likely has two minima during the orbital period of the companion~\citep{2010AstL...36..808G}. The inclined inner parts of the protoplanetary disk can screen its periphery from the radiation of the star. It was shown that there should be a shadow zone behind the inner region rising above the disk~\citet{2013AstL...39...26D}. At the same time, the image of the protoplanetary disk itself has a clearly defined horseshoe-shaped
region~\citep{2014AstL...40..334D,2015A&A...579A.110R}, which does not rotate with the disk. Similar results were obtained in~\citet{2018MNRAS.475.3201A,2019MNRAS.483.4221Z,2019MNRAS.484.4951N}.

Images of disks with horseshoe-shaped bright regions were obtained using the ALMA for the objects IRS48, HD 142527, AB Aur, HD 135344B~\citep{2021AJ....161...33V}. In addition, a quasi-stationary shadow on the disk of HD 135344B was discovered in~\citet{2017ApJ...849..143S}.

If the inclination angle of the planet's orbit is $\ge 60\degree$ then two symmetrical shadows may be visible on the image of the protoplanetary disk when observed in scattered light~\citep{2019MNRAS.483.4221Z}. A similar picture is also observed in images obtained in scattered light for the disks of the stars HD 142527~\citep{2015ApJ...798L..44M, 2017AJ....154...33A} and HD 100453~\citep{2017ApJ...838...62L,2017A&A...604L..10M}. It is assumed that the inclination of the inner disk relative to the outer one reaches $\sim 70\degree$. A substellar companion on an elongated orbit was discovered around
HD 142527~\citep{2016A&A...590A..90L,2019A&A...622A..96C}.

When observing in the submillimeter range, two symmetrical shadow regions can be observed even at smaller disk inclination angles~\citep{2015A&A...579A.110R}. Examples of objects with two shadow regions observed with ALMA are SR 21, J1604--2130, LkCa15, Sz91~\citep{2021AJ....161...33V}.

The question of the formation of a planet on an
orbit inclined relative to the disk remains open. Is the
tilt of the inner disk a consequence of the planet's
motion or was the planet itself formed from the matter
of the inclined inner disk? A number of studies have
shown that the planet's orbit tends to decrease its
eccentricity and tilt angle due to interaction with the
matter of the protoplanetary disk~\citep{2012ARA&A..50..211K,2013MNRAS.431.1320X}. Also, a review~\citet{2021AJ....161...33V} describes 38 objects that show asymmetries in the images of their disks. It is noted that only a few objects have evidence of a companion.

One of the mechanisms of distortion of the plane of
the protoplanetary disk is the capture of a cloudlet
from the remains of the protostellar cloud by a star. A
cloudlet falling onto a protoplanetary disk may be the
cause of the formation of an exoplanet on an inclined,
highly elongated or retrograde orbit~\citep{2011MNRAS.417.1817T}. The capture of a cloudlet from a protoplanetary disk and its accretion onto a star were modeled in the papers~\citet{2018MNRAS.475.2642K,2019A&A...628A..20D,2021A&A...656A.161K}. It
was shown that the fall of a cloudlet can lead to the
occurrence of an inclination of the outer part of the
disk relative to the inner, as well as to an elliptical
shape of the disk. In addition, in the case of a retrograde fall of a cloudlet, different parts of the disk can
move in opposite directions. For the first time, the
possibility of forming nested disks with opposite directions of movement due to the fall of matter from the
interstellar medium was shown in~\citet{2015A&A...573A...5V}. Calculations taking into account the magnetic field of the
disk and the cloudlet were carried out in~\citet{2022ApJ...941..154U}. The authors showed that if the size of the cloudlet is less
than or equal to the thickness of the disk, then the
magnetic field slows down the rotation of the falling
matter, but in the case of a sufficiently large cloudlet,
the matter can move at super-Keplerian speed.

Two-dimensional calculations~\citep{2005ApJ...633L.137V} shown that
clumps of matter with masses of several tens of Jupiter
masses can form in massive disks under the influence
of gravitational instability. They can accrete onto the
star, causing a burst of its accretion activity. In addition, such clumps can be ejected from the system into
interstellar space~\citep{2012ApJ...750...30B,2017A&A...608A.107V}. The possibility of the formation of dense clumps from the remains of a protostellar cloud was noted in~\citep{2018MNRAS.475.2642K}, and the accretion of such clumps can occur on the internal parts of the parent protostar ($\sim 1$~AU).

A collision of such a clump with the protoplanetary
disk of a protostar or another star in a dense star cluster
should lead to a distortion of the disk plane in the
region of its fall, which, in turn, may contribute to the
formation of a planet on an inclined orbit. The consequences of such a fall were studied in~\citet{2022ApJ...930..111D,2023ApJ...953...38D, GrigoryevDemidova_PCT2023}. 
It was assumed that the clump and the disk had reached thermal equilibrium. The problem of the decay of an arcshaped disturbance with an inclination of the initial velocity vector relative to the disk plane in a rotating disk medium was considered under the assumption of
an instantaneous and discrete fall of the gas stream.

As the cloudlet approaches the star, it can stretch into a stream of finite size. Similar structures have recently been discovered in ALMA images of several protoplanetary disks~\citep{2023ASPC..534..233P}. For example, a stream-like structure is observed in the young objects SU~Aur~\citep{2021ApJ...908L..25G} and DG~Tau~\citep{2022A&A...658A.104G}. Thus, the fall of dense gas flows onto a protoplanetary disk is not an exceptional event. At the same time it was shown in~\citet{2024MNRAS.528.6581H} that the stream of matter can be kept from dissipation by interstellar gas of higher temperature. The authors studied the fall of the stream before contact with the protoplanetary disk.

In this paper, we perform a three-dimensional simulation of the fall of a stream of matter onto the inner part of a protoplanetary disk and consider the dynamics of the reaction of the protoplanetary disk. A study of the physical properties and observational manifestations of such an event is carried out. In particular, the distortion of the inner region of the disk, the change in the accretion rate, and the variations in density along the line of sight are analyzed.

\section{MODEL AND METHOD}

\subsection{Basic Equations}

We will solve a system of non-stationary gas-dynamic equations describing the evolution of the gas flow around a young T Tau star using the  PLUTO\footnote{\href{http://plutocode.ph.unito.it/}{http://plutocode.ph.unito.it/}} \citep{PLUTOCODE} package in a spherical coordinate system $(R, \theta, \varphi): 144 \times 60 \times 144$ cells in the domain $[0.2; 107.2]$~AU $\times  [15; 165]\degree$ $\times [0; 360)\degree$.

Continuity equation
\begin{equation} \label{eq:masscons}
	\displaystyle \frac{\partial \rho}{\partial t} + \nabla \cdot \left( \rho \mathbf{v} \right)  =  0
\end{equation}
includes gas density $\rho$ and total velocity $\mathbf{v}$.
We will write the equation of gas motion taking into account
viscosity:
\begin{equation} \label{eq:navierstokes}	
	\displaystyle\frac{\partial (\rho \mathbf{v})}{\partial t} + \nabla \cdot \left( \rho \mathbf{v} \cdot \mathbf{v} -  p \hat{I} \right)^T =  - \rho \nabla \Phi + \nabla \cdot {\sf \Pi}(\nu).
\end{equation}
Here, $p$~is a gas pressure, $\hat{I}$~is unit matrix,  $\Phi = -GM_*/R$ gravitational potential created by a star ($G$~is gravitational constant, $M_* = 1 M_{\odot}$~is the star mass, $R$~is distance to the star, $M_{\odot}$~is Solar mass) and ${\sf \Pi}(\nu)$~is viscous stress tensor:
\begin{equation}
	{\sf \Pi}(\nu) = \nu_1 \left[ \nabla \mathbf{v} + (\nabla \mathbf{v})^T\right] + \left(\nu_2 - \frac{2}{3}\nu_1 \right)(\nabla \cdot \mathbf{v})\hat{I},
\end{equation}
$\nu_1$~is kinematic viscosity coefficient, $\nu_2$~is second viscosity (assumed to be equal to zero).

We will write the energy equation taking into account thermal conductivity:
\begin{equation} \label{eq:energycons}
	\displaystyle\frac{\partial (\varepsilon_t + \rho \Phi) }{\partial t} + \nabla \cdot \left[ \left(\varepsilon_t + p + \rho \Phi \right)\mathbf{v}   \right]  = \nabla \cdot (\mathbf{v} \cdot {\sf \Pi}(\nu)) + \nabla \cdot \mathbf{F}_c.
\end{equation}
Here, is the total energy density $\varepsilon_t = \rho \epsilon + \rho \mathbf{v}^2/2$ includes the specific internal energy of the gas $\epsilon$. Heat flow is determined by the thermal conductivity coefficient $\kappa$ and the gas temperature gradient $T$: $\mathbf{F}_c = \kappa \cdot \nabla T$.

To close the system of equations (\ref{eq:masscons}--\ref{eq:energycons}) the ideal gas equation of state is used: $p = \rho \epsilon (\gamma - 1)$, $\gamma = 7/5$.

It should be noted, this paper did not consider processes associated with radiation transfer (heating by UV radiation from hot regions of the star, cooling of dust). Also, the self-gravity of the disk was not taken into account: the gravitational potential $\Phi$ remained
unchanged during the calculations.

\subsection{Models of Kinematic Viscosity and Thermal Conductivity}

The velocity of the stream matter relative to the protoplanetary disk is very high, so mutual friction and heat exchange are inevitable during a collision. The stream matter will lose a significant fraction of kinetic energy due to such interaction, so the viscosity and thermal conductivity were calculated not only taking into account the turbulent terms, but in a more complex way.

All gas during the calculations is considered ideal and, within a separate cell, homogeneous. However, it is worth keeping in mind that, depending on the density and temperature, the gas can be in a different state (in particular, ionized), which determines its final viscosity and thermal conductivity. Each of transport coefficients consists of three components: for neutral gas, ionized, and a turbulent term. The formulas below are written in the CGS system.

The dynamic viscosity of neutral hydrogen was calculated according to the hard sphere model:

\begin{equation}
\mu_{neutral} = 1.016 \times \frac{5}{16 d^2} \sqrt{\frac{2 m_H k T}{\pi}},
\end{equation}
where $d = 2.9 \times 10^{-8}$~cm (diameter of a hydrogen molecule), $m_H$~is mass of hydrogen atom, $k_b$~is Boltzmann constant, $T$~is absolute temperature.

The viscosity of ionized plasma was determined based on Braginskii's formulas \cite{Braginskii1965en}: 
\begin{equation}
	\mu_{ion} = 0.96 \frac{k_b T}{\rho} \tau_i,
\end{equation}
where $\tau_i$~is a characteristic time of collisions between
ions in a fully ionized plasma..

Turbulent viscosity was specified according to the Shakura--Sunyaev model \cite{ShakuraSunyaev1973}:
\begin{equation}
	\mu_{turb} = \alpha \rho c_s H,
\end{equation}
where $\alpha = 0.001$, $c_s$~is a local speed of sound, and $H$~--- is a characteristic vertical scale of the disk (at the initial
moment of time it is determined by the formula~\ref{eq:H}, see below). In the calculations it was assumed that $H=\frac{c_s}{\Omega}$, where $\Omega$~is local Keplerian angular velocity. This relation is applicable in the case of vertical hydrostatic equilibrium, which is maintained until the stream substance touches the disk surface. Subsequently, when the initial plane of the disk is distorted,
a surface of maximum density can be identified, relative to which the average characteristic vertical scale of the disk can be estimated (by analogy with the distribution~(\ref{eq:rho}), see below: $\frac{\rho(r,\phi,z-z_0=H)}{\rho_{max}(r,\phi,z_0)}=e^\frac{1}{2}$ at a fixed $r$ and $\phi$). Calculations showed accordance in
order of magnitude between the characteristic vertical scale of the disk obtained in this way and the ratio of the speed of sound to the local Keplerian velocity during the entire simulation time. The deviation $\frac{c_s}{\Omega}$ from the value $H$ can be corrected by variations of the parameter $\alpha$ within the range $10^{-4}-10^{-2}$. Variations of $\alpha$ can be justified by physical processes occurring in the disk and influencing the transfer of angular
momentum~\citep[see, for example,][]{2008apsf.book.....H}, but not taken into account in this simulations.

The thermal conductivity of neutral hydrogen was also determined based on the hard sphere model:
\begin{equation}
	\kappa_{neutral} = \frac{1}{3} \rho c_v \lambda v_{therm},
\end{equation}
where $c_v$~is specific heat capacity of gas at constant volume, $\lambda$~is mean free path of a molecule (but not more than the size of the calculation cell), $v_{therm}$~is thermal velocity of gas.

The thermal conductivity of ionized hydrogen is primarily determined by electrons, so it can be calculated based on Braginskii's formulas:
\begin{equation}
	\kappa_{el} = 2 \times 10^{-8} T^{5/2}
\end{equation}

We will assume that the turbulent Prandtl number is equal to 1, then the turbulent thermal conductivity is determined on the basis of the known turbulent viscosity:
\begin{equation}
	\kappa_{turb} = c_p \mu_{turb},
\end{equation}
where $c_p$~is specific heat capacity of gas at constant
pressure.

The final viscosity and thermal conductivity are calculated as follows:
\begin{equation}
	\rho \nu_1 = (1 - x_{HI})\mu_{ion} + x_{HI}\mu_{neutral} + \mu_{turb}; \qquad \kappa = (1 - x_{HI})\kappa_{el} + x_{HI}\kappa_{neutral} + \kappa_{turb},
\end{equation}
where $x_{HI}$~is the fraction of neutral hydrogen, which was calculated based on the equilibrium of the processes of ionization and recombination of hydrogen in the cell.

The motivation for such a non-trivial model of
transport coefficients is as follows. In the disk corona,
especially near the star, laminar ion viscosity and electron thermal conductivity dominate, while in the disk, turbulent viscosity and thermal conductivity dominate. Therefore, it is necessary to create a sufficiently smooth and physical representation of the corresponding coefficients depending on the available macro parameters of the gas in each individual cell. In view of the use of formulas for a fully ionized plasma, the question of the degree of ionization of the gas naturally arises, which, in turn, allows us to talk about molecular transport coefficients.
	
Analysis of these coefficients showed that molecular viscosity is negligible compared to turbulent viscosity, which dominates in the gaseous ``cold'' disk, and ion viscosity, which predominates in the rarefied ``hot''ionized corona of the disk. However, in the thin
transition layer between the corona and the disk, its contribution to the total viscosity can be a fraction of a percent.
	
Molecular thermal conductivity is also negligibly small compared to turbulent in the plane of the disk, but becomes comparable to it when passing from the disk to the corona. In the corona itself, due to its ionization, the contribution of molecular thermal conductivity is zero. In addition, when the clump crosses the plane of the disk, active processes of heating and mixing of matter occur, thus one can expect the creation of conditions similar to those initially present in the region between the disk and the corona.

Thus, although the contribution of molecular viscosity and thermal conductivity is small, taking them into account allows us to create a more complete and accurate model for these transport coefficients within the framework of the gas-dynamic approximation.

\subsection{Initial and Boundary Conditions, Normalization Units}
All macroscopic parameters of gas are specified in a conventional system of code units. The main parameters of this system are listed in the Table~\ref{tab:dimensionalization}.

\begin{table}[!h]
	\caption{Normalization units} \label{tab:dimensionalization}
\bigskip
\begin{center}
		\begin{tabular}{|l|c|c|c|l|}
			\hline
			Parameter & Designation & Value  & Units, CGS  & Comment     \\[2pt] \hline
			Unit of mass     & $M_0$       & $1.98 \times 10^{33}$  & g          & $M_{\odot}$     \\
			Unit of length     & $L_0$       & $1.496 \times 10^{13}$ & cm         & $1$~AU        \\
			Unit of time   & $t_0$       & $3.16 \times 10^7$     & s          & $1$~yr         \\
			Unit of density & $\rho_0$    & $5.94 \times 10^{-7}$  & g/cm$^{3}$ & $M_0/L_0^3$     \\
			Unit of velocity  & $v_0$       & $4.74 \times 10^{5}$   & cm/s       & $2 \pi L_0/t_0$ \\
			\hline
		\end{tabular}
\end{center}
\end{table}

Same as in \cite{2022ApJ...930..111D} let's define the initial density distribution in the disk:

\begin{equation}\label{eq:rho}
	\rho(r, z, 0) = \frac{\Sigma_0}{\sqrt{2 \pi} H(r)} \frac{r_{in}}{r} e^{-\frac{z^2}{2 H^2(r)}}; \qquad \Sigma_0 = \frac{M_{disk}}{2 \pi r_{in}(r_{out} - r_{in})}.
\end{equation}
Here $r = R\sin \theta$~is cylindrical radius, $z$~is altitude
above the equatorial plane, $H(r)$~is characteristic half-thickness of the disk at this radius, $r_{in}$~is inner radius of the disk, $\Sigma_0$~is average surface density of the disk, $M_{disk}$~is initial mass of the disk (in all calculations it is set
equal to $0.01 M_{\odot}$).

The half-thickness of the disk depends on the temperature in the equatorial plane of the disk $T_{mid}$ at a given radius:
\begin{equation}\label{eq:H}
	H(r) = \sqrt{\frac{\kappa T_{mid}(r) r^3}{G M_\ast \mu m_H}}; \qquad T_{mid}(r) = \sqrt[4]{\frac{\Gamma}{4}}\sqrt{\frac{R_\ast}{r}} T_*,	
\end{equation}
where are the quantities $R_\ast$ and $T_\ast$ mean the radius and temperature of the star, the coefficient $\Gamma = 0.05$~\citep{1997ApJ...490..368C,2004A&A...421.1075D}, molar mass of gas in disk $\mu = 2.35$~\citep{1994A&A...286..149D}. 

The density in the cell is limited from below with the value $10^{-12} \rho_0$, and the entire region occupied by such a rarefied medium is considered a corona with a temperature $1.5 \times 10^6$~K. 

The velocity of each point on the disk is given based on the Keplerian approximation: $\mathbf{v} = (v_R, v_{\theta}, v_{\varphi}) = (0, 0, \sqrt{GM_*/R})$. In the case of calculations with retrograde fall of matter, the initial azimuthal velocity is replaced by a negative one: $v_{\varphi} \rightarrow -v_{\varphi}$. Thus, when the stream falls retrograde, the disk rotates clockwise in the plane $xy$ , and when it falls prograde, it rotates counterclockwise.

The left boundary condition for $R$ is set from considerations of equality of thermal turbulent flows, continuity of viscous turbulent flow and conservation of entropy. The values in the boundary cells are designated by the subscript $b$, without one is in the calculated cells close to the boundary:
\begin{equation}
	\kappa_{turb~b} \frac{\partial T_b}{\partial R} = -\kappa_{turb} \frac{\partial T}{\partial R}; \qquad \nu_{turb~b} \frac{\partial \rho_b \mathbf{v}_b}{\partial R} = \nu_{turb} \frac{\partial\rho \mathbf{v}}{\partial R}; 
	\qquad 
	\rho_b = \rho (T_b/T)^{2.5}. 
\end{equation}
in this case the component $v_{Rb}$ is always directed towards the star or zero.

The right boundary condition for $R$ corresponds to
the initial state of the matter in the disk. Free boundary conditions for $\theta$ on both sides and periodic boundary conditions for $\varphi$ are established.

\subsection{Initial Parameters of the Stream Matter}

At the initial moment of time, one calculation cell ($\sim 0.8$~AU) at a distance of $R = 20$~from the star is filled with a matter with a mass equal to that of Jupiter, and the density in it will be ($\rho\approx 1\cdot 10^{-9}$ g cm$^{-3}$). The initial temperature of this substance is $50$~K. The initial components of the gas velocity in this cell are set such that a material point with the coordinates of the center of this cell moves along a parabolic trajectory
with the longitude of the ascending node $\Omega = 45\degree$ relative to the star. Several variants of the stream and protoplanetary disk motion were considered, the parameters of which are listed in Table~\ref{tab:clump}. All angles are specified in degrees.

\begin{table}[!h]
	\caption{Initial parameters of the orbit of the falling stream substance. 
	} \label{tab:clump}
	\bigskip\begin{center}
		\begin{tabular}{|c|c|c|c|c|c|c|c|c|}
			\hline
			Parameter       & DI-45  & DI-r45 & DI-60  & DI-r60 & DI-30  & DI-r30 & SI-45 & SI-r45 \\ \hline
			$I,\degree$      & $45$   &  $45$  &  $60$  &  $60$  & $30$   &  $30$  & $45$  &  $45$  \\
			$\omega, \degree$ & $-110$ & $-110$ & $-110$ & $-110$ & $-110$ & $-110$ & $-30$ & $-30$  \\
			$q$, AU      & $5$    &  $5$   &  $5$   &  $5$   & $5$    &  $5$   & $7.4$ & $7.4$ \\
			\hline
		\end{tabular}	
		\end{center}
Designations of the parameters of the initial parabolic trajectory:  $I$~is inclination to the plane $xy$, $\omega$~is argument of pericenter and $q$~is pericentric distance. DI~is model with double intersection of the plane $xy$ by the initial parabolic trajectory, SI~is with single intersection. The ``r'' mark means the disk is rotating in the opposite direction (retrograde model).
\end{table}

The calculation parameters were selected so that the first intersection of the plane of the disk by the center of mass of the falling substance was approximately in the same place $(R, \theta, \varphi) \approx (7.4$~AU, $0\degree$, $225\degree$).

\section{RESULTS}

\subsection{Preliminary Statistical Analysis of the Motion
of Bodies on Parabolic Trajectories}

Before solving the gas-dynamic problem described above, it is useful to analyze the probability of a body flying toward the star from the outside crossing the plane of the disk.

It was assumed that the magnitude of the total
velocity of the falling matter at the initial moment of
time was equal to the escape velocity from the star, $V_{II}(R)=\sqrt{\frac{2GM_{*}}{R}}$, but directed towards the star.
During such a flyby, either a double or a single intersection of the disk plane is possible. Therefore, a study was made of the motion of the body along a parabola relative to the disk plane, limited by the radius $R=100$~AU. $10^6$ particles were placed in parabolic orbits
at a distance of $R=200$~AU. The orbital parameters
were set randomly: the pericentric distance ($q$) was in
the range $0.05-100$~AU, the inclination $0\degree \le I\le 180\degree$, argument ($\omega$) and longitude ($\Omega$) of the pericenter within $[0;360)\degree$.

For the estimation it was sufficient to consider that
the particles are gravitationally attracted only to the
star, and the influence of the disk (gas-dynamic drag,
disk gravity) was not taken into account. The integration of the orbits was performed using the Bulirsh--Stoer method~\citep{1980ina.book.....P}, the implementation of which is
described in~\citep{2022A&C....4100635D}. The calculation results were averaged over three variants of initial data.

Calculations showed that single disk plane crossings occur in $73\%$ of cases, and double ones in $27\%$. However, the number of orbits that cross the disk plane twice increases with decreasing argument of
pericenter $q$. Thus, for a fixed $q=5$~AU (other parameters correspond to those described above), double crossings account for $76\%$.

\subsection{Single Intersection of the Disk Plane}

Two variants of a single intersection of the disk
plane were considered. In the first case, the flow of
matter moved along a parabola in the same direction
as the disk matter, in the second case, the movement
was retrograde.

\begin{figure}[t!]
	\includegraphics[width=\columnwidth]{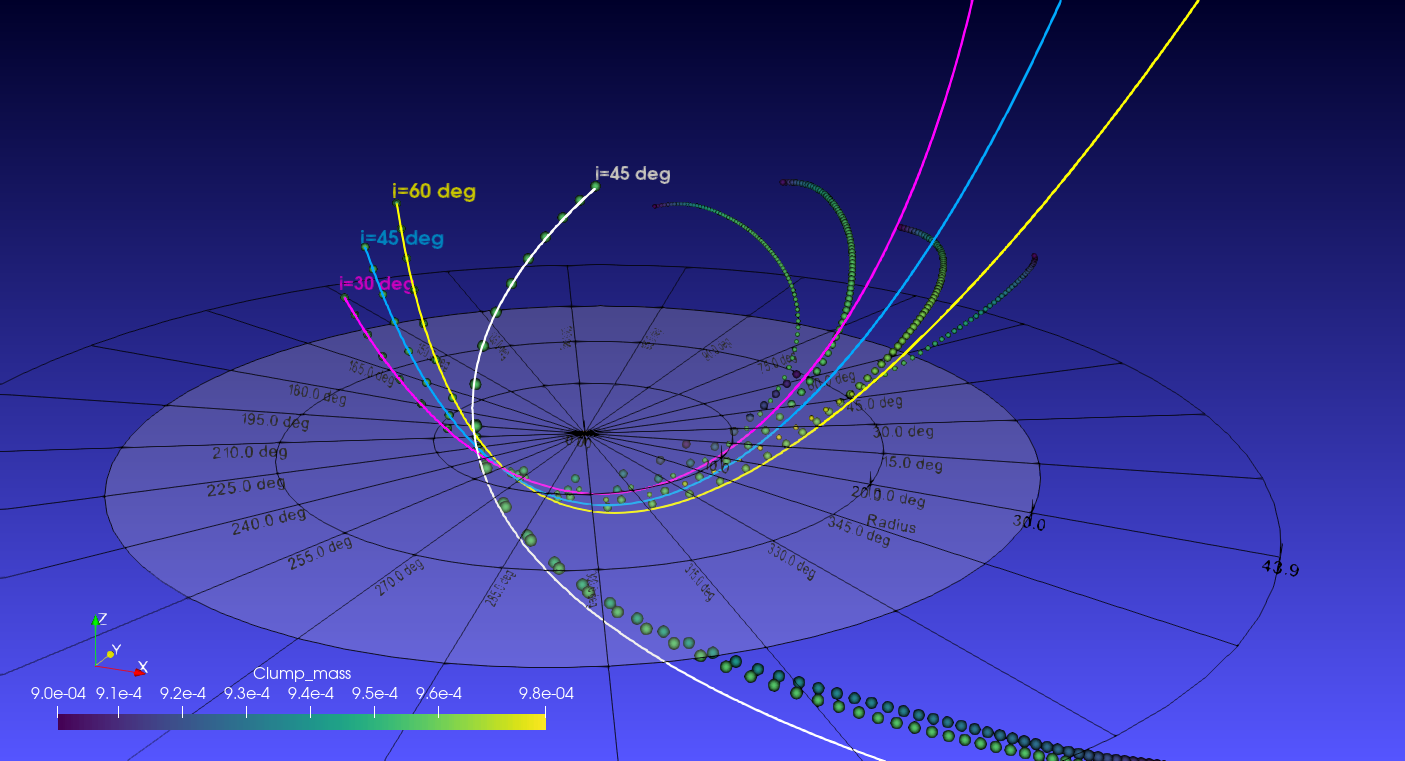}
\caption{Trajectories of the center of mass of the falling substance in the celestial-mechanical approximation (colored lines) and calculated taking into account gas-dynamic effects (colored balls). The plane of the disk is shown in translucent white, the inclination and azimuthal lines are indicated in degrees, the radial coordinate is in astronomical units. }
	\label{fig:pars}
\end{figure}

When moving toward the disk, the matter is stretched into a gas stream, which moves along the initial parabolic orbit until it touches the disk at the moment $t_{c1}$. Therefore, the distance $R_{c1}$ from the star at the moment the center of mass of the falling matter
crosses the plane of the disk is practically the same as
for the unperturbed orbit of a body moving along a parabola (Fig.~\ref{fig:pars} and Tab.~\ref{tab:t_and_R}). During the collision, the matter of the stream and the disk mix, and the orbit of
the matter changes, part of the matter of the stream is
captured by the gravity of the star, and the other, having passed through the disk, flies beyond the boundary of the domain. 

\begin{table}[!h]
	\caption{Moments of intersection of the disk plane by the center of mass of the falling matter ($t_i$, in years from the start of
calculations) and the distance to the star ($R_i$, in AU) at these moments for the calculated models} \label{tab:t_and_R}
	\bigskip \begin{center}
\begin{tabular}{|l|c|c|c|c|c|c|c|c|}
	\hline
	               & DI-45 & DI-r45 & DI-60 & DI-r60 & DI-30 & DI-r30 & SI-45 & SI-r45 \\ \hline
	$t_{1}$, years  &  6.7  &  6.7   &  6.9  &  6.9   &  6.6  &  6.6   &  9.3  &  9.3   \\
	$t_{c1}$, years &  6.9  &  6.9   &  7.1  &  7.1   &  6.7  &  6.8   &  9.5  &  9.5   \\
	$R_1$, AU    &  7.4  &  7.4   &  7.0  &  7.0   &  7.8  &  7.8   &  7.4  &  7.4   \\
	$R_{c1}$, AU &  7.3  &  7.3   &  6.9  &  6.9   &  7.7  &  7.5   &  7.2  &  7.1   \\ \hline
	$t_2$, years    & 14.9  &  14.9  & 16.2  &  16.2  & 14.0  &  14.0  &   -   &   -    \\
	$t_{c2}$, years & 17.0  &  14.6  & 19.5  &  17.0  & 15.7  &  13.4  &   -   &   -    \\
	$R_2$, AU    & 15.5  &  12.2  & 17.8  &  17.8  & 13.8  &  13.8  &   -   &   -    \\
	$R_{c2}$, AU & 15.8  &  11.3  & 18.6  &  14.7  & 14.1  &  8.9   &   -   &   -    \\ \hline
\end{tabular}
\end{center}
The numerical index corresponds to the intersection number. Index ``c'' is results obtained from gas-dynamic calculations, without this
index is in the celestial-mechanical approximation.
\end{table}

At the time of 50 years after the start of the calculations (when the interaction of the stream with the disk has already occurred, but the clump matter has not yet reached the outer boundary of the domain) in the SI-45 model the amount of clump matter accreted onto the star or moving in the calculation region with a velocity less than the local escape velocity is $0.70$ Jovian mass. In the retrograde case (SI-r45) at the same time within the domain $0.77$ Jovian mass are captured according to the same criterion.

\begin{figure}[t!]%
	\begin{center}
	\includegraphics[width=0.45\columnwidth]{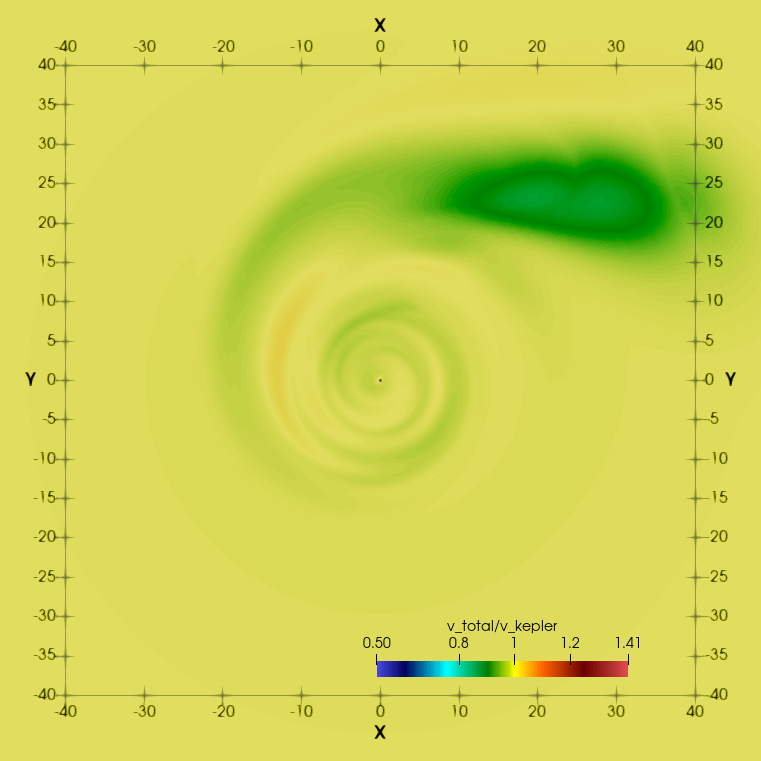}
	\includegraphics[width=0.45\columnwidth]{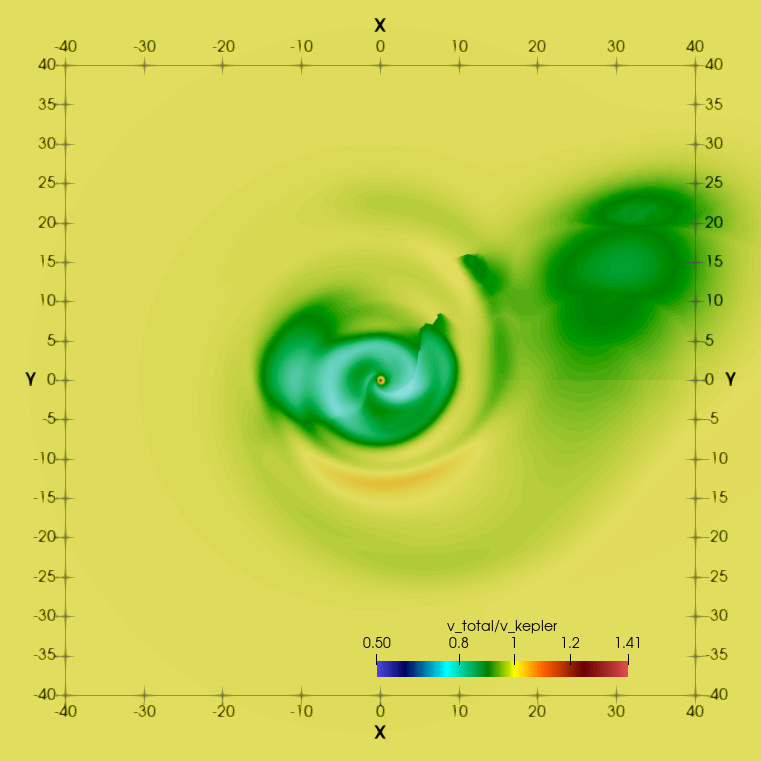} \\
	\includegraphics[width=0.45\columnwidth]{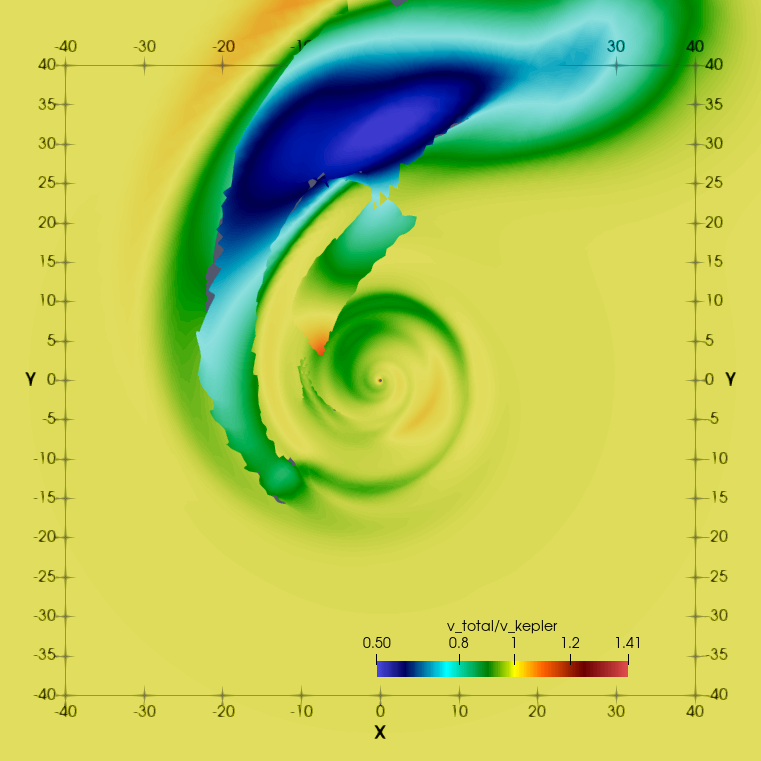}
	\includegraphics[width=0.45\columnwidth]{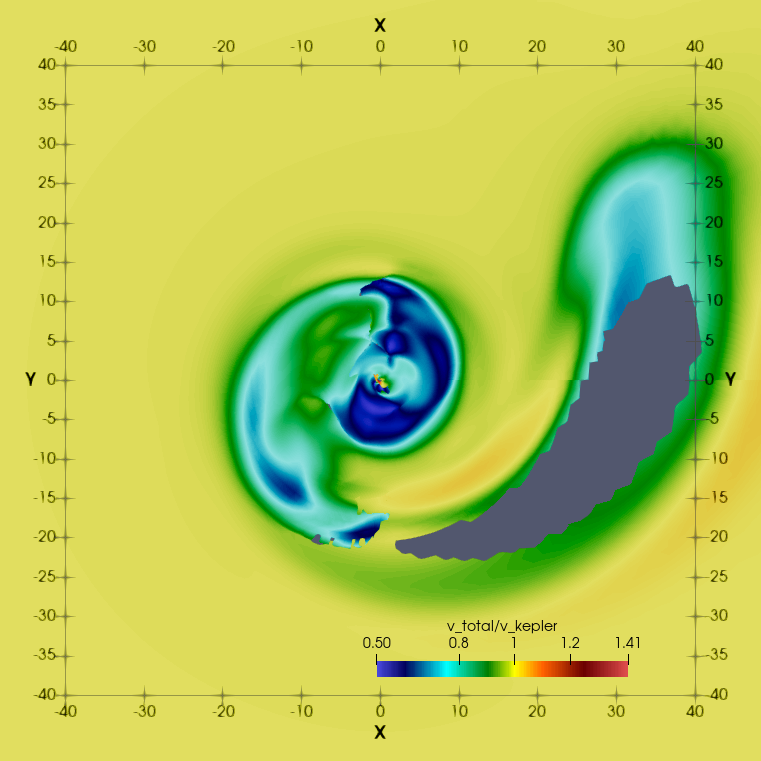}
	\end{center}
	\caption{Ratio of the total velocity $v_{total}$ to the local Keplerian velocity $v_k$ at a time of 50 years for four models (from left to right, top to bottom): SI-45, SI-r45, DI-45, DI-r45 on the surface of maximum density. X and Y are in AU.}
	\label{fig:v_tov_div_v_kep}
\end{figure}

After the stream passes through the disk, it is distorted both horizontally and vertically, and the velocity of the matter changes. Figure~\ref{fig:v_tov_div_v_kep} shows the ratio of
the local velocity of the matter on the surface of maximum density to the Keplerian velocity at a given distance for different models. After the collision, a single-arm spiral density wave propagates inward and outward from the disk. Regions arise in which the current
velocity is less than the local Keplerian velocity. In the
case of retrograde fall, a large region of the disk has a
velocity half of the Keplerian velocity. 

\begin{figure}[t!]
	\includegraphics[width=1.0\columnwidth]{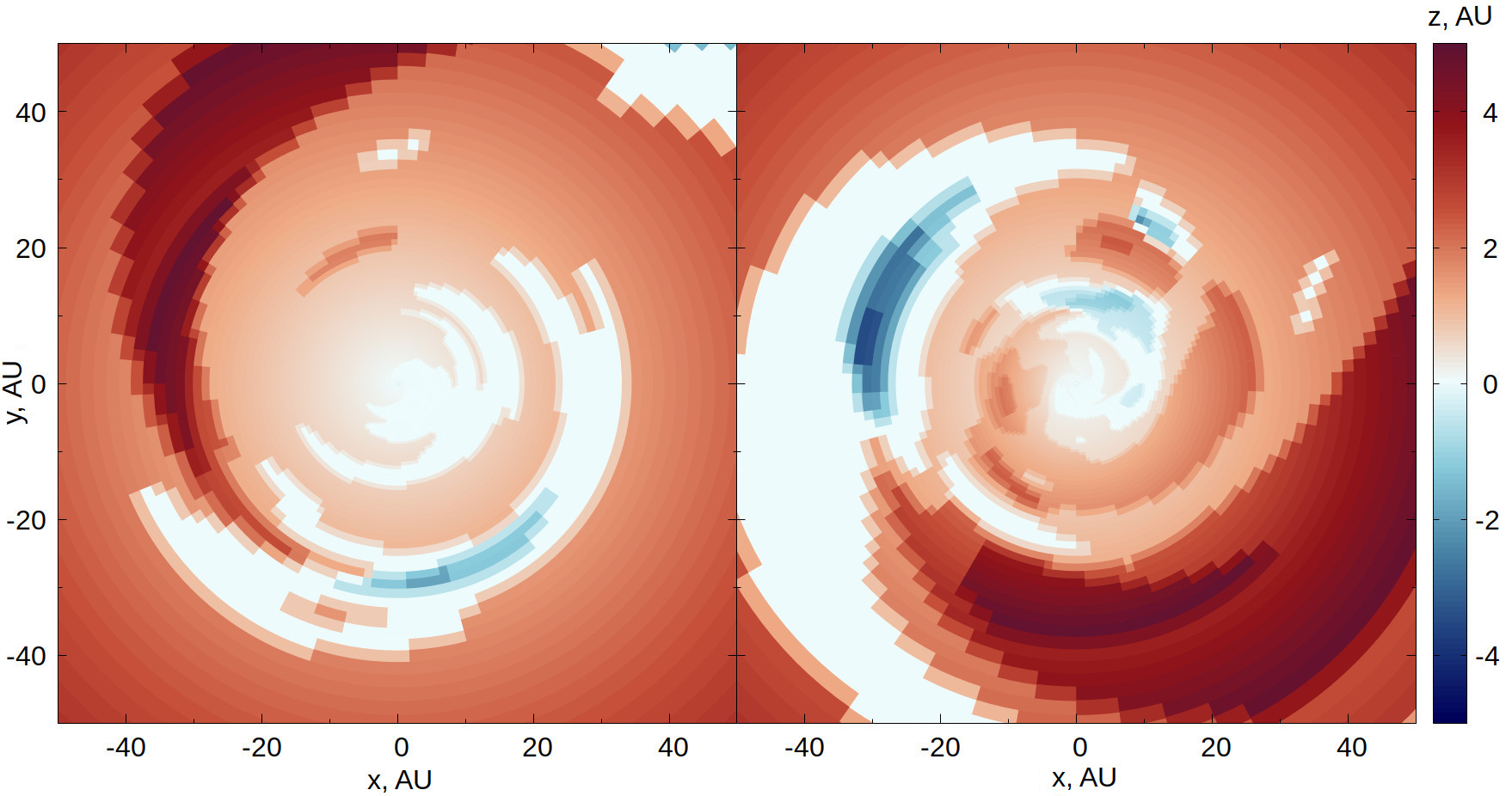}
\caption{Values of the coordinate $z$, at which the density of the disk matter is maximum in cells at a fixed $(r, \varphi)$ at the time of 150 years from the start of the calculations. The prograde case (SI-45) is shown on the left, and the retrograde case (SI-r45) is shown on the right.}
	\label{fig:rZ-SP}
\end{figure}

The initial plane of the disk is also distorted significantly. In Fig.~\ref{fig:rZ-SP} the values of the vertical coordinate $z$, are shown in color, at which the maximum density in
a cell with fixed $(r, \varphi)$ in cylindrical coordinates is
achieved. It is evident that 150 years after the start of
the calculations the disk is strongly distorted in the vertical direction. The periphery of the disk is raised
relative to its initial plane. It is evident that the spiral
wave propagating along the disk is also raised relative
to its initial plane. On the inner part of the disk it is
possible to distinguish areas of the surface rising above
the initial plane of the disk and going under the disk,
however it is not possible to distinguish the inclined
inner disk.

Due to the presence of a spiral wave caused by a
single intersection of the plane of the disk by the
stream matter, it can be concluded that if a second
intersection occurs from the opposite side in the
opposite direction, then an internal inclined disk can
be expected to appear, since two spiral waves with
oppositely directed humps are formed, diverging in
different directions. In addition, due to the presence of
two points of interaction between the clump and the
disk, an increase in the fraction of the clump mass
captured by the disk can also be expected.

\subsection{Double Intersection of the Disk Plane}

During its motion, the stream substance moves along a parabolic trajectory until the first contact, which occurs at a time $t_{c1}$ at a distance $R_{c1}$ from the star. Here, the stream matter is violently mixed with the disk matter, and a portion of the matter is heated and carried away by the disk. After this, the center of mass
of the fallen matter reaches the pericenter of the orbit and then intersects the disk plane for the second time at a time $t_{c2}$ at a distance $R_{c2}$. The values of these parameters for each model are given in Table~\ref{tab:t_and_R}. All these times and distances differ from those calculated in the celestial-mechanical approximation, since the latter does not assume the presence of drag and interaction of matter. However, the corresponding trajectories are close to each other almost until the moment of the second contact with the original disk plane (Fig.~\ref{fig:pars}).

As expected, several spiral waves propagate across
the disk when the initial plane of the disk is crossed
twice. The velocity of the matter decreases due to the
mutual reduction of the angular momenta of the
stream and the disk. Fifty years after the start of the
calculations, the velocities of the matter near the star
are noticeably less than the local Keplerian ones
(Fig.~\ref{fig:v_tov_div_v_kep}). At the same time, in the case of retrograde fall, there are regions where the total velocity of the gas
is less than $65$\% of the Keplerian velocity. The presence of such velocities should lead to an acceleration of the fall of matter onto the star, as previously shown in~\citet{2022ApJ...930..111D}.

At the time of 50 years after the start of the calculations, it is possible to estimate the total mass of the clump matter, which moves slower than the local free fall velocity and also accreted onto the star. For the DI-60 and DI-r60 models, it is equal to $0.84$ and $0.85$ Jovian mass, respectively; for the DI-45 and DI-r45 models, it is $0.87$ and $0.89$ , respectively; for the DI-30 and DI-r30 models, it is $0.90$ and $0.91$, respectively.

\begin{figure}[t!]
	\includegraphics[width=\columnwidth]{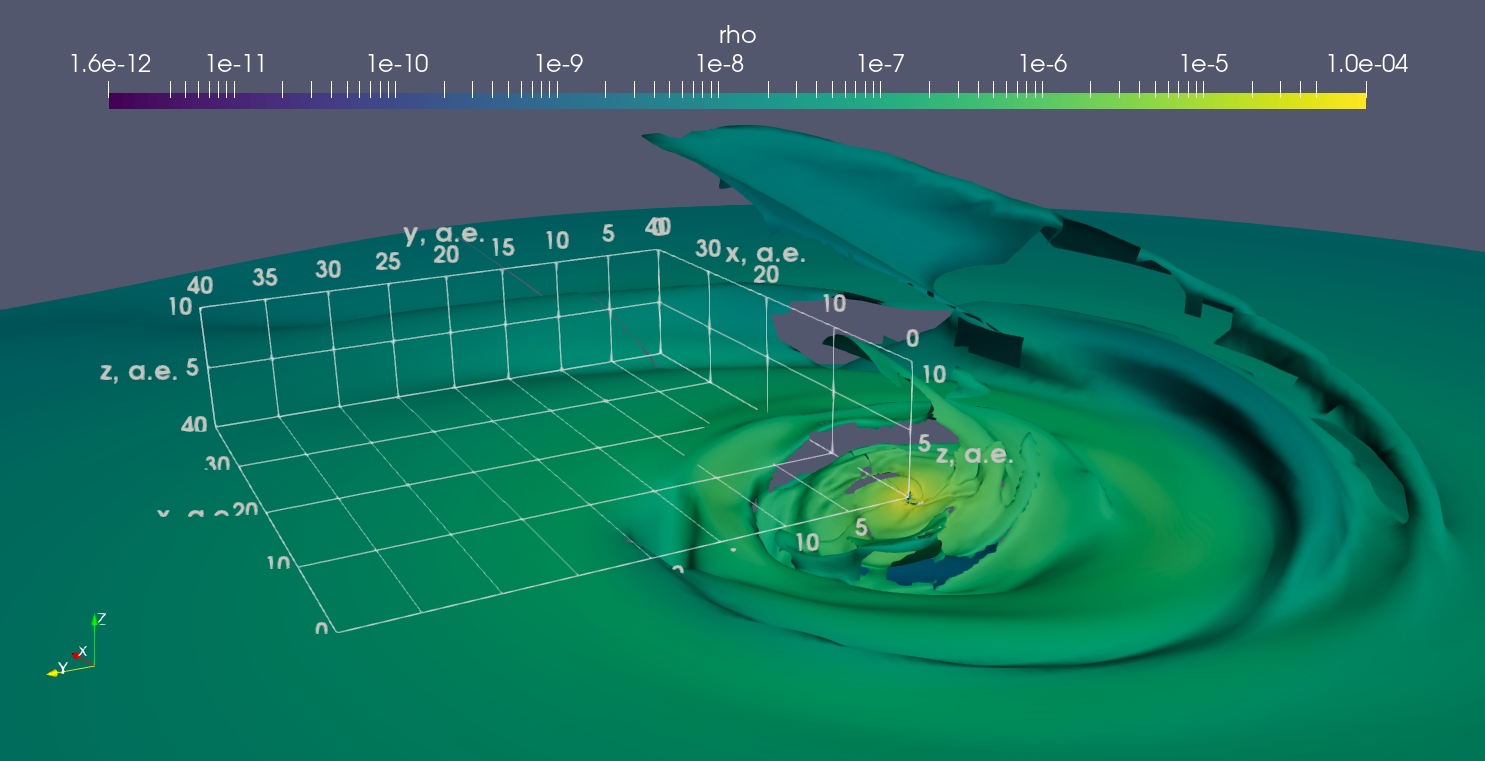}
\caption{Surface of maximum density at the time $90$ years relative to the start of calculations in the DI-r45 model. The color shows
the density in units $\rho_0$. The coordinate parallelepiped is shown for scale. X, Y and Z are in AU.}
	\label{fig:incl-DPr45rho-3D}
\end{figure}

\begin{figure}[t!]
	\includegraphics[width=1.0\columnwidth]{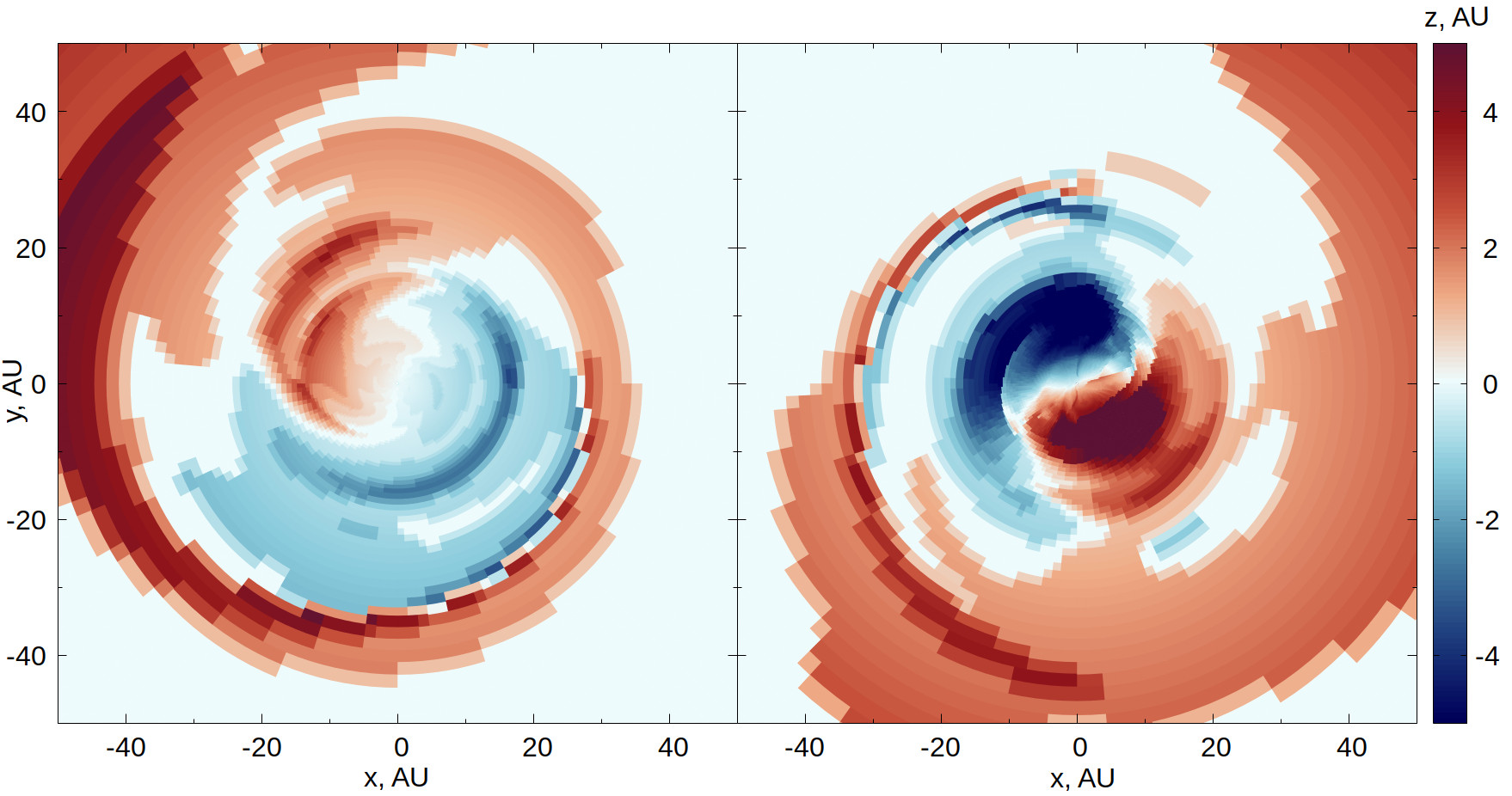}
\caption{The same as in Fig.~\ref{fig:rZ-SP} for the case of a parabolic orbit crossing the plane of the disk twice.}
	\label{fig:rZ-DP}
\end{figure}

Figure~\ref{fig:incl-DPr45rho-3D} shows the density distribution along the surface of maximum density in the disk in the DI-r45 model at the time $90$~years, which illustrates well the distortion of the disk plane. It can be seen that the distortion is substantially nonuniform, there are spiral waves of density and rarefaction, the presence of waves is also noticeable in Fig.~\ref{fig:rZ-DP}, and the distorted region extends more than 40 AU from the star. It is clearly
seen that the central inner part of the disk is significantly inclined relative to the periphery, which we will call the inner inclined disk (warp).

The values of the vertical coordinate $z$, at which the
maximum density is achieved are shown in Fig.~\ref{fig:rZ-DP} at
the time of $150$ years. In this case, quasi-symmetric
distortions of the disk plane in its inner part can be distinguished, half of which is shifted to the positive part of the axis $z$ , and the other half to the negative. In this case, for the prograde and retrograde fall of matter, the pattern of displacement of the region of increased density is diametrically opposite. This effect is due to the fact that the disk rotates counterclockwise in the prograde case, and clockwise in the retrograde case. In Fig.~\ref{fig:rZ-DP}, the stream moves from the observer, crosses the picture plane and goes under the disk, and then exits it towards the observer. Accordingly, the regions of the disk in which the substance first goes under the disk and then rises are shifted in the direction of the Keplerian rotation of the disk: counterclockwise in the case of prograde fall, and clockwise in the retrograde one.

\begin{figure}[t!]
	\includegraphics[width=0.45\columnwidth]{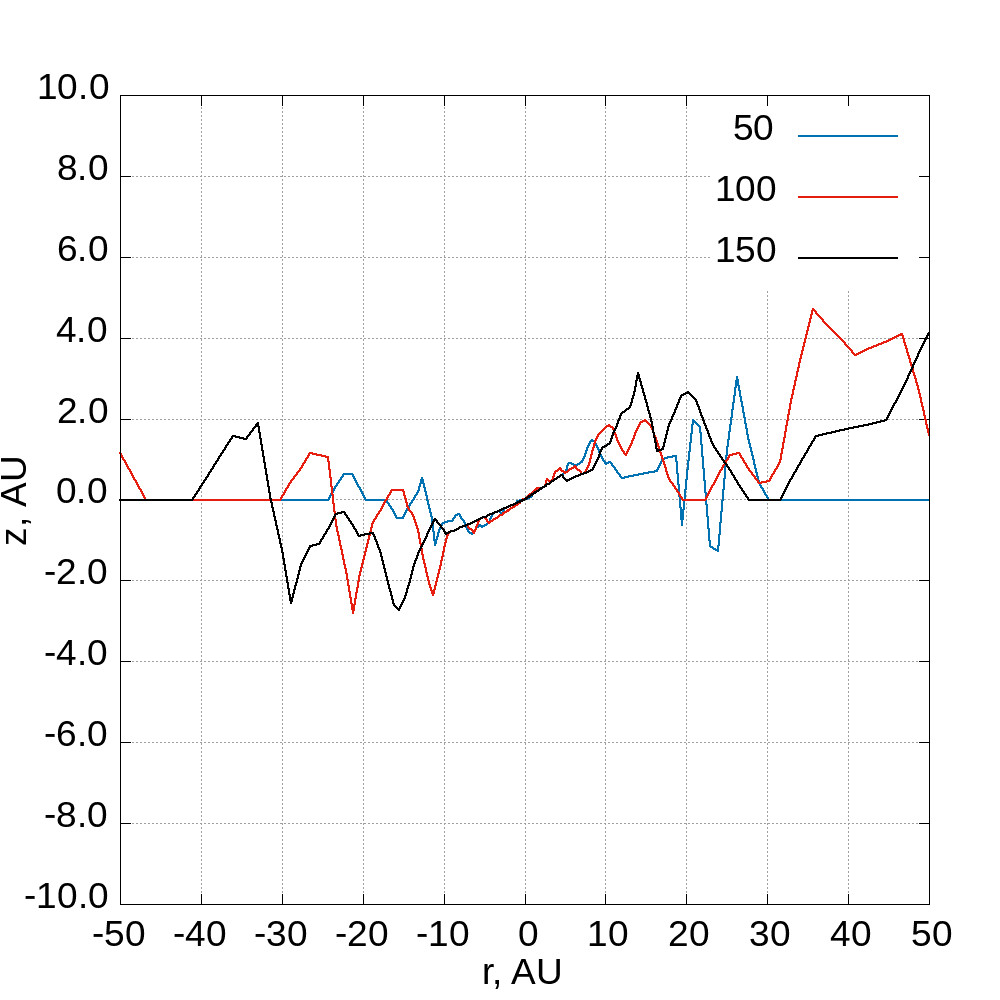}
	\includegraphics[width=0.45\columnwidth]{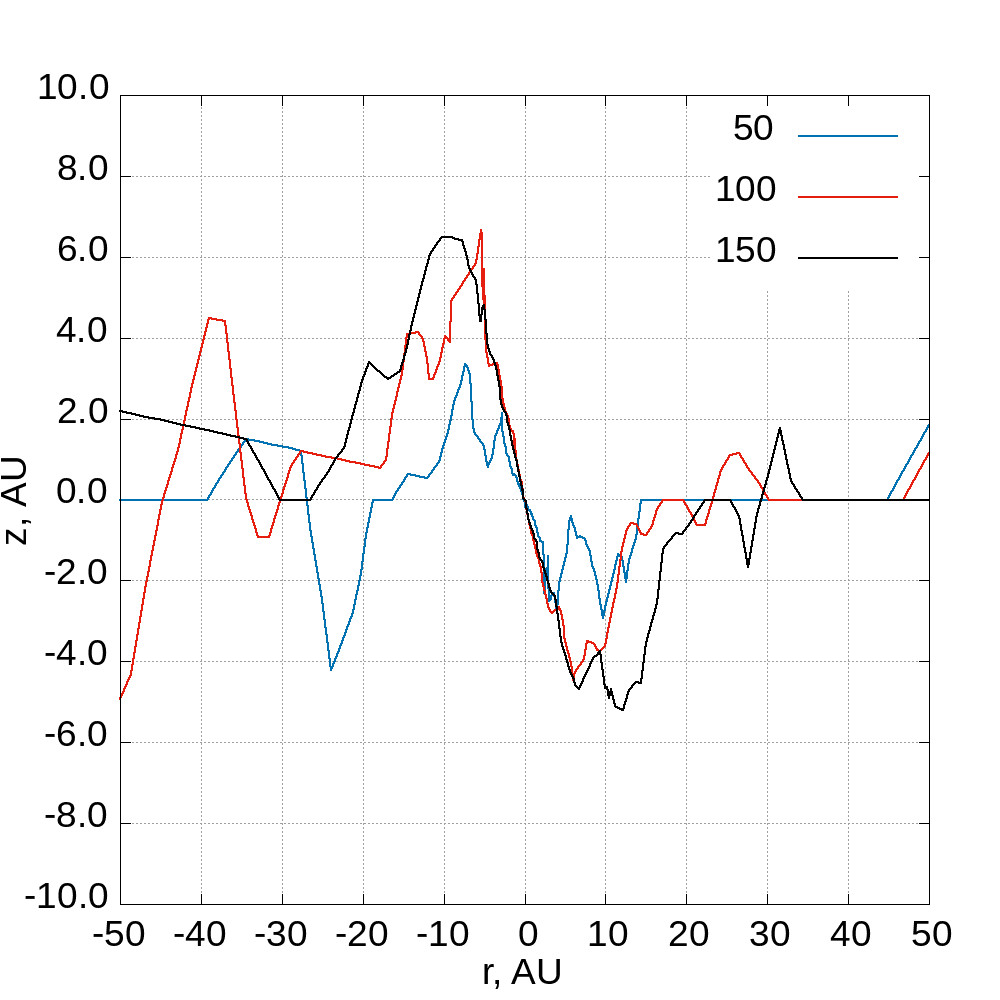} \\
\caption{Values of the coordinate $z$, at which the density of the disk matter is maximum depending on the distance ($r$) at fixed $\varphi=135\degree$ ($r>0$) and $\varphi=315\degree$ ($r<0$) at the moments of time 50 years (blue line), 100 years (red line), and 150 years (black line) from the beginning of the calculations. The case of prograde fall is shown on the left, and retrograde fall on the right.}
	\label{fig:inclSD}
\end{figure}

The position of the regions of maximum density along two azimuthal angles $\varphi=135\degree$ and $\varphi=315\degree$ (since the tilt of the inner disk is maximum for the DI-45 and DI-r45 models near this direction) is shown in Fig.~\ref{fig:inclSD} for three time moments of 50, 100, and 150 years from the start of the calculations. It is evident that in the retrograde case the inner disk is noticeably inclined relative to the periphery, whereas in the case of a prograde
flyby this inclination is less pronounced. Thus, the retrograde fall of the stream through the protoplanetary disk leads to the formation of a distinct inclined inner disk with a noticeable inclination relative to the initial plane of the disk. In addition, in Fig.~\ref{fig:inclSD} one can dynamically trace the propagation of the spiral density wave across the disk: the wave-like distortion of the
vertical plane of the disk shifts over time from the region of collision with the stream to the periphery.

The initial inclination of the orbit of the infalling
material relative to the plane of the disk affects the inclination angle of the inner disk. Table~\ref{tab:params_of_warp} lists the
parameters of the inner inclined disk at three points in time for all models with double intersection. The boundary of the inner disk is an diverging spiral wave on the surface of maximum density, the crest of which is located on different sides of the plane $xy$, which
makes it difficult to reliably determine the size. Therefore, the size of the inner disk $R_{warp}$ was determined by
the average radius of this diverging spiral wave. Within $R_{warp}$ from the star the average normal to the surface of
maximum density was calculated, after which the
angle $i$ between this normal and the axis $z$ (also known
as the angle between the planes of the disks) and the azimuth of the normal $\Phi$, which is $90^\circ$ from the line of
nodes (i.e., the line along which the planes of the outer
and inner disks intersect), were determined.

\begin{table}[!h]
		\caption{Parameters of the internal inclined disk for models with double intersection of the disk plane} \label{tab:params_of_warp}
		\bigskip
		\begin{center}
		\begin{tabular}{|c|c|c|c|c|c|c|c|}
			\hline
			 t, years  &     Parameter     & DI-45 & DI-r45 & DI-60 & DI-r60 & DI-30 & DI-r30 \\ \hline
			          &   $i,^\circ$    &   8   &   22   &  10   &   17   &   5   &   23   \\
			   50     &  $\Phi,^\circ$  &  309  &  107   &  308  &  129   &  311  &  109   \\
			          & $R_{warp}$, AU &   7   &   7    &   7   &   7    &  11   &   7    \\ \hline
			          &   $i,^\circ$    &   9   &   20   &  11   &   26   &   7   &   16   \\
			   100    &  $\Phi,^\circ$  &  310  &  103   &  312  &  106   &  312  &   71   \\
			          & $R_{warp}$,AU &  11   &   9    &  10   &   9    &  12   &   12   \\ \hline
			          &   $i,^\circ$    &   8   &   25   &   8   &   16   &   6   &   10   \\
			   150    &  $\Phi,^\circ$  &  314  &  106   &  328  &  121   &  317  &   77   \\
			          & $R_{warp}$, AU &  15   &   13   &  13   &   13   &  15   &   15   \\ \hline
		\end{tabular}
	\end{center}
		
Here, $i$~is angle to the plane $xy$ (initial plane of the disk), $\Phi$~is azimuth of the normal, $R_{warp}$~is maximum size at times of 50, 100, 150 years.		
\end{table}

It is evident that over time the inclined region expands along the radius of the disk. In this case, in the models of prograde stream fall, the angle of inclination of the inner disk changes little or decreases, whereas in the case of retrograde fall, the angle of
inclination of the disk mainly increases with time
(a more detailed analysis of the warp evolution in the
DI-r45 model is described in~\cite{GrigoryevDemidova_SVMO2024}), except for the DI-r30 model, where the inclination is reduced. In the
case of a stream substance falling at an angle $30^\circ$  to the
plane of the disk, it is quite difficult to isolate the
region of the inner disk due to the large number of
crests and cavities on the surface of maximum density,
especially at times close to the moment of the fall. The
greater the initial angle of inclination of the orbit of the
falling substance, the less disturbed and more symmetrical the inclined inner disk becomes.

Separately, we note that in the case of a prograde
fall of the stream, the plane of the inner disk is oriented in azimuth approximately in the same direction as the plane of the initial orbit of the stream (the angle $\Phi$ differs from $225^\circ$ approximately by $\sim 90^\circ$). In the case of a retrograde fall, such agreement is not observed.

\subsection{Disk Mass and Accretion Rate}

\begin{figure}[t!]
	\includegraphics[width=0.45\columnwidth]{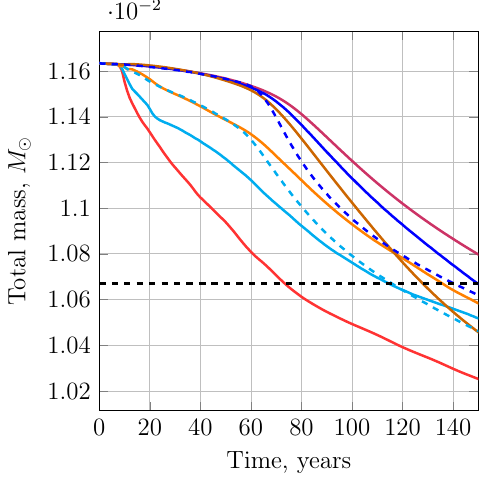}
	\includegraphics[width=0.45\columnwidth]{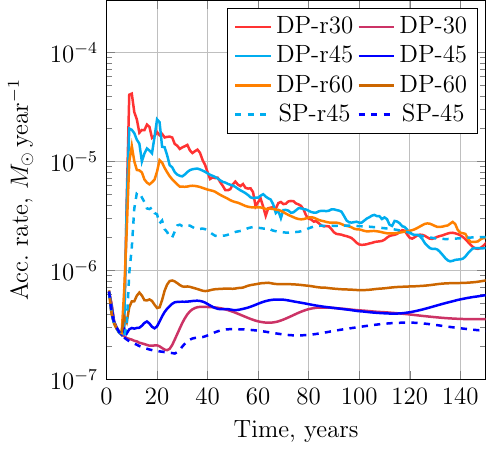}
\caption{Left: change in the mass of the matter in the computational domain as a function of time. Solid lines correspond to models
with double intersection of the disk plane, dashed lines is with single intersection. The black dashed line on the left graph shows
the initial mass of the disk (without the stream matter). Right: the accretion rate into a sphere of radius $R=0.2$~AU depending
on the model. The legend is common to both graphs.}
	\label{fig:mass}
\end{figure}

The total mass of the matter in the computational domain decreases due to accretion onto the star and the outflow of the matter beyond the computational domain (Fig.~\ref{fig:mass} on the left). In models with a stream falling in the same direction as the disk rotation
(SI-45, DI-30, DI-45, and DI-60), the mass of the matter changes with time almost equally throughout the entire computation time. In the case of a retrograde flyby, the mass of the matter begins to decrease
noticeably after about 9 years; this process is associated with a sharp increase in the rate of accretion onto the star. The total mass becomes equal to the initial mass of the disk after 40--80 years from the start of the computations.

In addition, the rate of accretion onto the star (the amount of matter passing through the minimum boundary $R$ per unit of time) changes in the system; it can significantly increase the overall luminosity of the system, thereby affecting the shape of the light curve.

The accretion rate at the initial moment of time is $\dot{M}\approx 3\cdot 10^{-7}M_\odot$/year. In models with a prograde
stream fall, the accretion rate changes slightly after the
collision and remains at the same level. With retrograde fall, the accretion rate increases sharply. In the  case of a single intersection of the disk plane (SI-r45), it increases by about 20 times in 4 years. Whereas, in the case of a double intersection of the disk plane by the orbit, the accretion rate increases by about
100 times in 2 years (Fig.~\ref{fig:mass} on the right), then, on
average, over 50 years the accretion rate decreases by
about 3 times. Thus, in the case of a retrograde stream
fall along an orbit that crosses the disk plane twice, an
FU Ori-like outburst of luminosity is possible due to
the rapid increase in the accretion rate~\citep{1977ApJ...217..693H, 2013MNRAS.434...38K, 2021ApJ...917...80S}.

\subsection{Opacity of the Disk}
Since the distribution of the protoplanetary disk matter is noticeably distorted after the stream of matter falls on the disk, the amount of dust in the line of sight also changes. Since submicron and micron-sized dust particles make the main contribution to extinction in the visible range of the spectrum, the column densities were calculated only for such fine dust. In order to calculate the amount of fine dust in the line of sight, the density of matter in the disk was integrated
along the radius for fixed values of $\theta$ and $\varphi$. It was
assumed that the total mass of fine dust in the entire disk is $10^{-5} M_{\odot}$. With the mass of the gas disk $M_g=0.01M_\odot$ it was assumed that the dust to gas mass ratio is 1:100 , as on average in the interstellar medium, and the amount of fine dust is $10\%$ of the total amount of dust in the disk. An increase in the
mass of fine dust proportionally increases the column density and, accordingly, the opacity. 

\begin{figure}[t!]
	\includegraphics[width=\columnwidth]{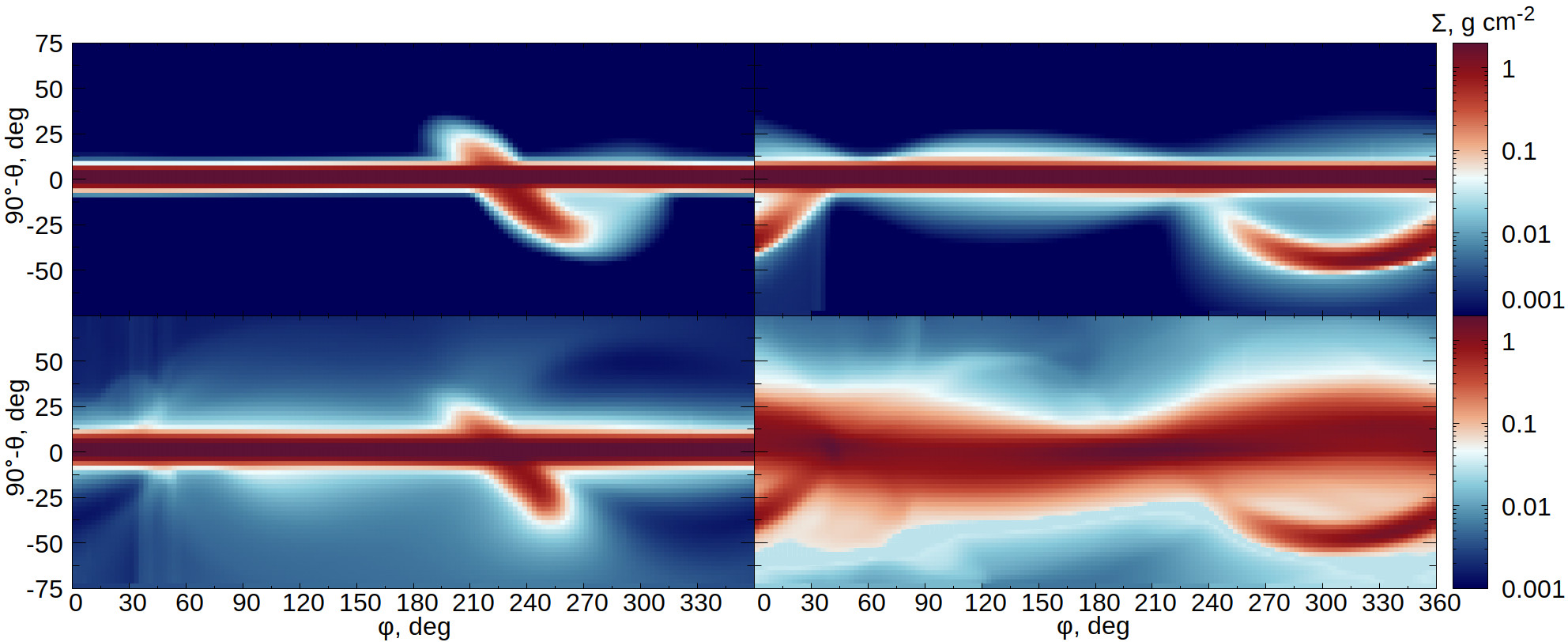}
\caption{The column density of the substance is shown in color depending on the direction of the line of sight in azimuth ($\varphi$) and inclination ($\theta$) at a time of 10 years. On the left are models in which the orbit of the falling substance intersects the plane of the disk once (SI-45 from above and SI-r45 from below), on the right is a double intersection (DI-45 from above and DI-r45 from
below).}
	\label{fig:sig10}
\end{figure}

\begin{figure}[t!]
	\includegraphics[width=\columnwidth]{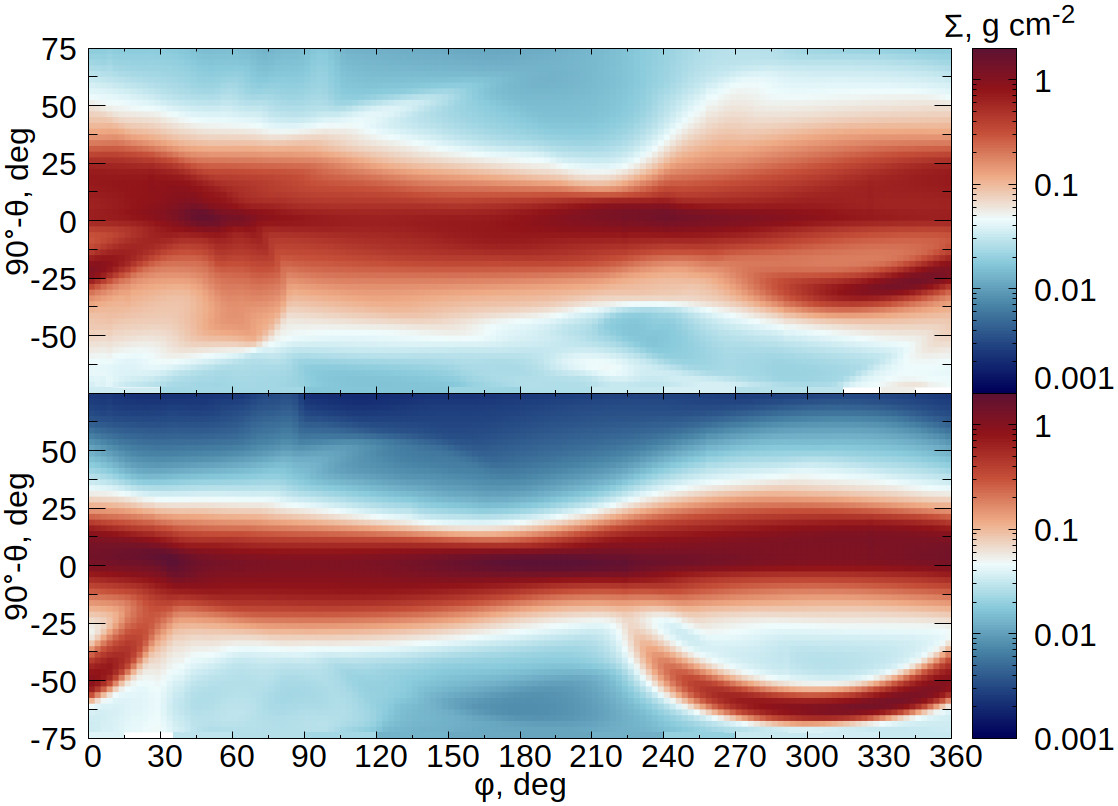}
\caption{The same as in Fig.~\ref{fig:sig10} for double intersection of the plane of the disk at angles $I=30^\circ$ (top) and $I=60^\circ$(bottom).}
	\label{fig:sigI}
\end{figure}

Figure~\ref{fig:sig10} shows the behavior of the dust density on
the line of sight at a time of 10 years depending on the
direction of the line of sight for all values of $\varphi$ and $\theta$ in cells. It is evident that during retrograde fall the disk
matter rises above the initial plane higher than in the
prograde one, which leads to a noticeable increase in
the density on the lines of sight intersecting the disk
plane at a large angle. In this case, in the model in
which the initial orbit of the stream intersects the disk plane twice, a significant density on the line of sight
can be observed at an angle $> 50^\circ$to the disk plane. It
is interesting to note that with a decrease in the angle
of inclination of the initial orbit of the falling matter,
the disk matter rises higher above the initial plane
(Fig.~\ref{fig:sigI}). This is apparently due to the fact that the falling matter travels a longer path in the disk, thereby
perturbing it more strongly.

\begin{figure}[t!]
 \includegraphics[width=\columnwidth]{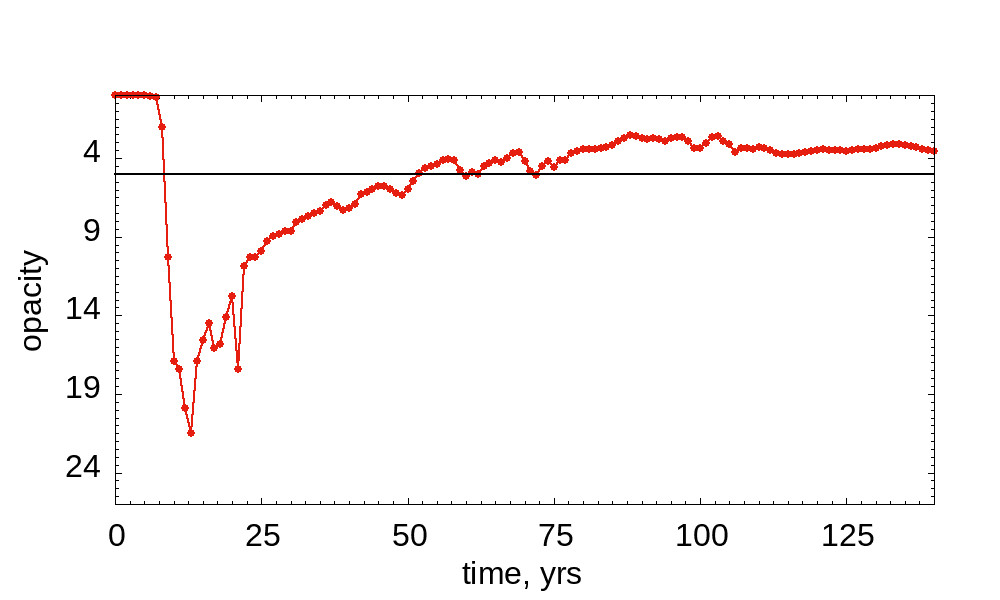}
\caption{Optical thickness on the line of sight in the direction $\varphi=90^\circ$ and $\theta=45^\circ$ in the band V (red line with dots). The black line shows the value $\tau=5$ for the DI-r45 model.} 
	\label{fig:tau}
\end{figure}

To calculate the optical thickness of fine dust
trough the line of sight, the column density must be
multiplied by the absorption coefficient. It can be
taken as $250$ cm$^2$/g for the Johnson photometric system band V~\citep{2000A&A...364..633N}. The optical thickness for the selected
direction $\phi=90^\circ$ and $\theta=45^\circ$ is shown in Fig.~\ref{fig:tau} for the DI-r45 model. It is evident that at the time of $\sim 10$~years after the start of the calculations, the optical thickness grows rapidly and reaches its maximum
within a few years. Then, a smooth decrease in the
optical thickness occurs. A qualitatively similar picture
is observed for other directions of the line of sight.

If the change in brightness $I$ were due only to
absorption, the change in stellar magnitude would
correspond to the value $\tau$: $\Delta m =-2.5*\lg(\frac{I\cdot e^{-\tau}}{I})=2.5\tau \lg(e)\approx \tau$. However, the decrease in the brightness of a star during an eclipse by a region on the disk
is limited by the scattered light from the entire disk as
a whole. Usually, its share is several percent of the
radiation of the star. Therefore, for example, the total
brightness cannot decrease by more than $\sim 5^m$, if the
scattered light is $1\%$ of the radiation of the star. Thus,
when converting the optical thickness shown in Fig.~\ref{fig:tau}, into stellar magnitudes, a sharp drop in brightness will
be observed on the light curve, the minimum will last
for $\sim 30$~years, and then a smooth growing to a bright
state will be observed. The duration of the minimum
depends on the chosen direction of the line of sight
and can range from several tens to hundreds of years.

\section{DISCUSSION}

Calculations have shown evidence that a significant increase in the accretion rate by two orders of magnitude is due to the appearance in the immediate vicinity of the star of a noticeable amount of matter
whose velocities are several times less than the local Keplerian velocity. Such conditions are realized in the case of a retrograde fall of the stream along a trajectory that crosses the initial plane of the disk twice at distances $\leq \sim 10$~AU. When matter falls on more distant parts of the disk, the effect is likely to be less pronounced. For example, as shown in~\citet{2018MNRAS.475.2642K,2019A&A...628A..20D,2021A&A...656A.161K}, in the case of a cloudlet falling onto the periphery of the disk, the luminosity outburst is an order of magnitude smaller than that observed in objects like FU Ori, and the accretion rate does not exceed $10^{-9} M_\odot yr^{-1}$. The
authors note that such an event could trigger the development of gravitational instability in the disk, which could subsequently lead to a significant increase in the rate of transfer of angular momentum to the star.

Another external mechanism that could trigger an increase in the accretion rate is the passage of another star close to the protoplanetary disk~\citep{2008A&A...492..735P,2010MNRAS.402.1349F}. Calculations taking into account radiative transfer showed
good agreement with the observed increase in the luminosity of FU~Ori-type in amplitude and time characteristics~\citep{2022MNRAS.510L..37B}. It is noted that the pericentric distance between the stars should be $\leq \sim 20$~AU. Moreover, the smaller the pericentric distance, the faster the accretion rate increases to its maximum. The presence of a protoplanetary disk near the incoming star increases the maximum accretion rate several times, and also allows to reproduce the rate of luminosity drop after the outburst~\citep{2022MNRAS.517.4436B}. However, as shown in~\citet{2023ARep...67.1401S}, a flyby of a star with a pericentric distance of $500$~AU can trigger the development of thermal instability in the inner part of the disk, with the subsequent launch of magneto-rotational instability, which subsequently leads to a delayed ($\sim 1000$~years after the approach) rise in the accretion rate by two orders of magnitude.

After the collision of the gas stream with the disk matter, the gas can rise high above the initial plane of the disk, which in turn leads to a sharp ($\sim 1$~yr) increase in the optical thickness on the line of sight by several times in the case of large angles of its inclination with a smooth subsequent decrease in the optical thickness. In this case, the scattered light limits the decrease in
brightness, so the light curve should show a sharp drop in radiation intensity by several stellar magnitudes, then a prolonged minimum, and then an exit to a bright state. A similar picture is observed in the light curve of two stars: CQ~Tau~\citep{2023Ap.....66..235G} and V1184~Tau~\citep{2015A&A...582A.113S}. The first object's brightness dropped by $3^m$ over several months, then the star remained near minimum brightness for about 80 years, and in the last year it began to brighten. In addition, this object has a tilted inner disk~\citep{2004ApJ...613.1049E,2008A&A...488..565C}. The second object experienced a brightness drop of $5^m$, then a minimum of 10 years, and then the star returned to its initial bright state within four years. In addition, the star AA Tau also experienced a brightness drop of $4^m.5$ in 2011~\citep{2013A&A...557A..77B} and for 12 years the star is being in a state of minimum brightness~\citep{2021AJ....161...61C}.  

An important issue of this study is the question of the survival of clump of matter until the moment of collision with the protoplanetary disk. In the case of the ejection of a clump formed in the process of gravitational instability in another protoplanetary disk~\citep{2017A&A...608A.107V}, it must travel thousands of AU before colliding.

In the problem under consideration, at the initial moment of time, a matter with a mass of $1$ Jupiter mass and a temperature of $50$~K fills a cell of size $\sim 0.8$~AU, the average density of the matter in this case is $1\cdot 10^{-9}$~g cm$^{-3}$. A spherical clump of such density and such temperature has a critical mass of $0.13$ Jovian mass. This means that the model clump presented in the paper should collapse due to self-gravity. The non-isothermality of the sphere, its rotation, and the non-uniform composition of the matter can prevent compression. This issue requires a separate study, which is beyond the scope of this paper. It should be noted that in the paper the initial clump is a numerical approximation for studying the response of the protoplanetary disk to the fall of a finite-size gas stream.

In~\citet{1991MNRAS.249..584C} formulas are given for estimating the probability of a star colliding with the protoplanetary disk of
another star. If these formulas are applied to the considered model of a clump colliding with a protoplanetary disk, then the probability of such an event can be estimated. In a collision with a disk near a star with a mass $1~M_\odot$ at a distance of $7.4$~AU from it during the
disk's lifetime ($\sim 10^6$ years), the probability is one
event per $8.8 \times 10^3/(Nf)$, where $N$ is the number of
clumps ejected by one disk, and $f$ is probability of survival of clumps. Thus, if $N>1$, $f\approx 1$ such an event is not extremely rare. 

In the case of a fall of matter from the remains of a protostellar cloud, clumps can form in the immediate vicinity of the protoplanetary disk and, as shown in~\citep{2018MNRAS.475.2642K}, accretion of such clumps can occur onto the inner parts of the parent protostellar system ($\sim 1$~AU). The authors note that episodes of late accretion are a common phenomenon for all calculated models of protostars. Thus, this mechanism of formation of clumps and their subsequent fall onto the protoplanetary disk in the form of a stream is most probable.

\newpage
\section{CONCLUSIONS}

In~\citet{2022ApJ...930..111D} the disintegration of a small mass gas clump in a rotating medium of a protoplanetary disk was investigated in the approximation of instantaneous occurrence of inhomogeneity as a result of the fall of matter from the external medium onto the disk
periphery. It was shown that the clump's matter, carried away by the disk's rotation, is first stretched into a vortex-like structure, then into a single-arm spiral, and then, depending on the initial velocity of the clump, a horseshoe-shaped asymmetry, a ring structure, or a double-arm spiral is formed. In the next paper~\citet{2023ApJ...953...38D} the case of a discrete fall of a massive clump
onto a gas disk near a star was considered. Evidence was obtained that, with a significant loss of kinetic energy by the matter of the fallen gas stream upon collision with the disk near the star, an inclined inner disk is formed. Since the perturbation in the disk at the initial moment of time is gravitationally bound to the star
and has velocities less than the Keplerian velocity, as well as a non-zero vertical velocity component, a second intersection of the unperturbed disk by the perturbation matter takes place. This apparently contributes to the formation of an inclined inner disk. It was also shown that if the disk contains a large region of matter
with velocities of $60\%$ of the Keplerian velocity, a sharp increase in the accretion rate is possible, resembling an FU~Ori-type outburst over time, which was additionally verified in \cite{GrigoryevDemidova_PCT2023}. 

In this work we consider the case of a continuous fall of matter onto a gas disk in the form of a finite-size stream. Unlike previous papers, where calculations were performed using the smooth particle hydrodynamic method (SPH), in this paper we used the finite-volume method. The results of these calculations confirmed the conclusions of previous papers. It turned out that the greatest losses of kinetic energy of the stream of matter occur in models in which the trajectory of the center of mass of the falling matter is retrograde and intersects the initial plane of the disk twice. Thus, it is in this case that the conditions described in~\citet{2023ApJ...953...38D} can be realized. It has been shown that at different initial angles of inclination of the initial orbit of the falling matter, an inclined inner disk is formed. Therefore, it can be assumed that the fall of the stream of matter can contribute to the formation of a planet
on an orbit inclined relative to the plane of the disk (and, accordingly, the plane of the stellar equator) near the star. This is what distinguishes this model from those described in~\citet{2018MNRAS.475.2642K,2019A&A...628A..20D,2021A&A...656A.161K}  where the periphery of the disk is tilted relative to its internal parts.

In addition, during the retrograde fall of the stream, a large region of matter is formed in the disk, the velocities of which are less than $65$\% of the Keplerian velocity at the same distance. Such matter begins to fall onto the star at an accelerated rate, thereby
increasing the accretion rate by two orders of magnitude, and then a smooth decrease in the accretion rate occurs during the relaxation of the disk. In this case, an FU~Ori-like outburst of luminosity can be
expected. It should be noted that the turbulent viscosity changes insignificantly during the collision of the stream with the disk matter. The initial mass of the clump, necessary to increase the accretion rate by two orders of magnitude, is 3 times smaller in this paper, and the collision distance is 3 times greater than that
obtained in~\citet{2023ApJ...953...38D}. This is due to the fact that the velocities of matter during a collision with a retrogradely
falling stream in the considered model are distributed
in a range of values, which includes values that are less
than $60$\% of the Keplerian velocity, and are not set equal for the entire disturbance, as in the previous paper.

Thus, the results of the numerical simulation carried out in this paper allow us to confidently link two observed features of protoplanetary disks: FU~Ori-type flares and the formation of internal inclined disks within the framework of one idea: the fall of a finite-
sized gas stream onto a protoplanetary disk.

The amount of stream matter captured by the disk or falling onto the star was analyzed. It can be concluded that the smaller the angle of inclination of the clump orbit, the longer the clump matter interacts
with the disk, and therefore the greater the percentage of clump matter remains in the system. As expected, in the retrograde case the percentage of captured matter is higher and reaches more than $90\%$ at the angle of inclination of the stream orbit of $30\degree$ in the case of double intersection of the disk plane than in the case
of prograde fall. In addition, the fraction of captured mass is noticeably greater for double intersection of the disk plane by the stream orbit than for a single one.

In this paper it was shown for the first time that after the collision of the stream substance with the disk the gas can rise high above the initial plane, and one can expect the gas to entrain fine dust. In such a case a sharp increase in the optical thickness (during the
year) in the directions $-50\degree<\theta<50\degree$, can be
observed, and then a smooth decrease in absorption. It should be noted that at the same time the luminosity outburst due to accretion will also affect the changes in the brightness of the star. However, as shown in~\citep{2024AL...GD}, an increase in the accretion rate also provokes an increase in the disk wind, which in turn will also screen
the star from the observer. Thus, all three effects: the rise of matter above the disk, the increase in the accretion rate, and the increase in the disk wind will probably be reflected in the shape of the light curve (including at large angles of inclination of the line of sight to the plane of the disk). First, there will be a sharp drop
in brightness by several stellar magnitudes, then a stay in the minimum, which can last for decades, and then a smooth growing to a bright state.

It should be noted that radiation transfer was not taken into account in these calculations. The calculation results showed that after the collision of the stream substance with the disk, the gas is heated to
several thousand Kelvins in the collision region due to adiabatic compression. Then the heated region is stretched into a spiral by the Keplerian rotation of the disk and spreads to the periphery of the disk, heating it. By the time of 150 years, the spiral structure is
located within a radius of $45$~AU. Thus, heat transfer occurs at a rate of $\sim 35$~AU per $\sim 140$~years. In~\citet{2017A&A...605A..30M} it was shown that the characteristic time of thermal relaxation of the disk matter with a mass $0.01M_\odot$ under the action of radiative processes at a distance of $7.5$~AU is $t_{relax}\approx 1/(10\Omega)\approx 0.3$~years and decreases with increasing distance and temperature. Thus, the disk should be effectively cooled by radiation. Then, as noted in~\citet{2024AL...GD}, the flash of accretion activity of the star, caused by the fall of the clump onto the disk, should also be accompanied by an increase in infrared radiation from the disturbed region of the disk.

The distribution of matter at the moment of time immediately after the collision is primarily determined by the energy of the falling stream. Subsequently, with an increase in temperature, the effective half-thickness of the disk increases. To study this process in calculating the long-term dynamics of gas in the protoplanetary disk after a collision with an external gas stream, a more detailed calculation of the heat balance is necessary. Such calculations are planned to be per
formed in a future.

\begin{acknowledgments}
The calculations were carried out using the resources of the Joint SuperComputer Center of the Russian Academy of Sciences, Branch of Federal State Institution ``Scientific Research Institute for System Analysis of the Russian Academy of Sciences''~\citep{2019LG..40..1835}.

The authors express their gratitude to the reviewer for
quality comments and interesting ideas that complement
the content of the paper.
\end{acknowledgments}

\bibliographystyle{rusnat}
\bibliography{maikbibl}
\end{document}